\documentclass[pra]{revtex4-2}
\usepackage{eurosym}
\usepackage{amssymb}
\usepackage{graphicx}
\usepackage{verbatim}
\usepackage[normalem]{ulem}
\usepackage{color}
\usepackage{amsmath}

\begin{document}

\title{Quantum-mechanical wave functions in singular potentials: linear and
nonlinear states}
\author{Hidetsugu Sakaguchi}
\affiliation{Interdisciplinary Graduate School of Engineering Sciences,
Kyushu University, Kasuga, Fukuoka 816-8580, Japan}
\author{Boris A. Malomed$^*$$^{1,2}$}
\affiliation{$^*$The corresponding author; e-mail: malomed@tauex.tau.ac.il\\
$^1$Department of Physical Electronics, School of Electrical and Computer Engineering,
Faculty of Engineering, Tel Aviv University, Tel Aviv University, Tel Aviv
69978, Israel\\
$^2$Instituto de Alta Investigaci\'{o}n, Universidad de Tarapac\'{a}, Casilla 7D, Arica, Chile}

\begin{abstract}
It is known that the attractive singular potential $\sim -1/r^{2}$ gives
rise to the critical quantum collapse in the framework of the
three-dimensional (3D) linear Schr\"{o}dinger equation. This article
summarizes theoretical results which demonstrates suppression of the
collapse ,caused by this singular potential, and the creation of the
otherwise missing ground state (GS) in a 3D gas of bosonic particles,
carrying an electric dipole moment, which are pulled to the central electric
charge, with repulsive contact interactions between the particles. In the
mean-fie;ld approximation, the repulsive interactions are represented by the
cubic term in the respective Gross-Pitaevskii (GP) equation. In addition to
the GS, excited states with angular momen tum are briefly considered too.
Another topic considered in the article is 1D and 2D bound states in the
linear-Schr\"{o}dinger and GP equations with the repulsive potential which
demonstrates a singularity at $r\rightarrow \infty $. A very recent result
that such a potential, growing faster than $-r^{2}$, produces a full
spectrum of counter-intuitive normalizable (localized) bound states. The
article puts forward perspectives for further studies of linear and
nonlinear bound states existing under the action of the potentials with the
singularity at $r\rightarrow 0$ or $r\rightarrow \infty $.
\end{abstract}

\maketitle


\textbf{Acronyms}: BEC -- Bose-Einstein condensate; BIC -- bound state in
continuum; GPE -- Gross-Pitaevskii equation; GS -- ground state; HOP --
harmonic-oscillator potential; ISP -- inverse-square potential; MFA --
mean-field approximation; QC -- quantum collapse; QD -- quantum droplet; VK
-- Vakhitov-Kolokolov (stability criterion); vNW -- von Neumann-Wigner (BIC
solution)



\section{Introduction and the models}

\subsection{The attractive inverse-square potential (ISP) with the
singularity at $r\rightarrow 0$}

The motion of quantum particles under the action of singular potentials may
be dramatically different from classical motion under the action of the same
potentials. A well-known example, which is a included in standard textbooks
on quantum mechanics, is provided by the three-dimensional (3D) Schr\"{o}%
dinger equation with the inverse-square potential (ISP) of pulling to the
center,
\begin{equation}
U(r)=-\frac{U_{0}}{2r^{2}},~U_{0}>0  \label{U}
\end{equation}%
\cite{LL}. The strength of the potential, $U_{0}$, may be eliminated from\
the corresponding classical equation of motion for the particle's position, $%
d^{2}\mathbf{r/}dt^{2}=U_{0}\mathbf{r/}r^{4}$, by means of the
straightforward obvious rescaling, $t\equiv \tilde{t}/\sqrt{U_{0}}$,
However, the invariance with respect to the variation of $U_{0}$ is not
maintained by the 3D Schr\"{o}dinger equation with the same potential, for
wave function $\Psi \left( \mathbf{r},t\right) $,%
\begin{equation}
i\frac{\partial \Psi }{\partial t}=-\frac{1}{2}\nabla ^{2}\Psi -\frac{U_{0}}{%
2r^{2}}\Psi   \label{Schr}
\end{equation}%
(written in the dimensional form), in which $U_{0}$ cannot be eliminated by
rescaling. This difference between the motion of the classical particle and
its counterpart in quantum mechanics is commonly known as the \emph{quantum
anomaly}, one of those which occur in diverse physical settings \cite%
{anomaly,anomaly2,superselection,Olshanii,HS1,Trib,ODell,anomaly-new}.

It is well known too that, if the external harmonic-oscillator potential
(HOP),%
\begin{equation}
U_{\mathrm{trap}}=\frac{1}{2}\Omega ^{2}r^{2},  \label{trap}
\end{equation}%
is added to the Schr\"{o}dinger equation (\ref{Schr}) in 3D, the interplay
of ISP and HOP produces a set of stationary wave functions, starting from
the ground state (GS), which are singular at $r\rightarrow 0$, but,
nevertheless, meaningful, as their norm,
\begin{equation}
N=\int \left\vert \Psi (\mathbf{r})\right\vert ^{2}d\mathbf{r,}  \label{N}
\end{equation}%
converges, provided that the ISP\ strength does not exceed a specific
critical value,
\begin{equation}
U_{0}<\left( U_{0}\right) _{\mathrm{cr}}^{(\mathrm{3D})}=1/4  \label{1/4}
\end{equation}%
(it is written in the notation adopted in Eq. (\ref{Schr})), while the \emph{%
quantum collapse} (QC, alias \textquotedblleft fall onto the center" \cite%
{LL}) at $U_{0}>1/4$ \cite{anomaly,anomaly2,HS1}. In this case, the GS does
not exist, while the evolution of the time-dependent wave function, governed
by Eq. (\ref{Schr}) with $U_{0}>1/4$, leads to the catastrophic compression
of the corresponding density distribution, $\left\vert \Psi (r)\right\vert
^{2}$, into a region of a vanishingly small radius, $r_{0}(t)\rightarrow 0$
at $t\rightarrow \infty $ (see a detailed analysis of the collapse dynamics
in Ref. \cite{Trib}). The situation is still more challenging in the 2D
version of Eq. (\ref{Schr}), as it leads to QC at any value $U_{0}>0$.

The ISP can be realized physically as proposed in Ref. \cite{HS1}. In that
case, quantum (bosonic) particles are realized as small polar molecules
(e.g., LiCs \cite{LiCs} or KRb \cite{KRb}) carrying an electric dipole
moment, $\mathbf{d}$. The particles are pulled to a point-like electric
charge, $Q$, placed at the center, which induces the electric field, $%
\mathbf{E}=Q\mathbf{r}/r^{3}$. To minimize the interaction energy, the
particle's dipole moment is locked to the local field, so that $\mathbf{d/}d=%
\mathrm{sgn}(Q)\left( \mathbf{r}/r\right) $. The corresponding electrostatic
potential of the particle is $U(r)=-\mathbf{d}\cdot \mathbf{E}$, producing
ISP (\ref{U}) with
\begin{equation}
U_{0}=2|Q|d.  \label{U0}
\end{equation}%
The same physical realization may be implemented for the 2D version of the
Schr\"{o}dinger equation (\ref{Schr}). Another realization of the 2D case is
possible in a gas of magnetically polarizable atoms, whose effective
magnetic moment is induced by the magnetic field of an electric current
(electron beam) piercing the 2D setting in the perpendicular direction \cite%
{HS1}.

A solution to the QC problem (i.e., replacing the collapsing dynamics by a
physically meaningful regular behavior) was proposed in Ref. \cite{HS1}:
replacing the linear Schr\"{o}dinger equation \ref{Schr} by its nonlinear
counterpart, in the form of the Gross-Pitaevskii equation (GPE), which is
also written in the dimensionless form:%
\begin{equation}
i\frac{\partial \Psi }{\partial t}=-\frac{1}{2}\left( \nabla ^{2}+\frac{U_{0}%
}{r^{2}}\right) \Psi +\left\vert \Psi \right\vert ^{2}\Psi .  \label{GPE}
\end{equation}%
In the framework of standard mean-field approximation (MFA) \cite{BEC}, Eq. (%
\ref{GPE}) is a model of the Bose-Einstein condensate (BEC) of the
electric-dipolar particles which are attracted to the central charge, while
the cubic term in Eq. (\ref{GPE}) accounts for the MFA effect of repulsive
collisions between the particles. It was demonstrated that the cubic
self-repulsive term in the GPE is sufficient to completely suppress the QC,
giving rise to a physically meaningful GS solution for all values of $U_{0}$%
, including those which \emph{exceed the critical value} (\ref{1/4}) \cite%
{HS1}.

It is relevant to mention that, in addition to the collisional (contact) MFA
nonlinearity, represented by the self-repulsive cubic term in Eq. (\ref{GPE}%
), the electrostatic interaction between the electric dipole moments of the
particles produces an additional contribution to the GPE. In the framework
of MFA the additional term amounts to the same cubic one, with a rescaled
coefficient in front of it \cite{HS1}.

\subsection{The expulsive potential with the singularity at $r\rightarrow
\infty $}

While the QC suppression by the nonlinearity in the framework of GPE (\ref%
{GPE}) with the singular ISP \cite{HS1}, and various extensions of this
setting \cite{HS2,HS3,GEA,Viskol,HS4}, were investigated in detail, another
challenging problem is offered by linear and nonlinear Schr\"{o}dinger
equations with the expulsive potential which features a singularity at $%
r\rightarrow \infty $. This problem was addressed very recently in Ref. \cite%
{Academia}. The basic model is provided by the 1D GPE,%
\begin{equation}
i\frac{\partial \Psi }{\partial t}=-\frac{1}{2}\frac{\partial ^{2}\Psi }{%
\partial x^{2}}-\frac{1}{2}x^{2\gamma }\Psi +g|\Psi |^{2}\Psi ,
\label{1DNLS}
\end{equation}%
with $\gamma >0$, and its 2D counterpart, written in terms of the polar
coordinates $\left( r,\phi \right) $:%
\begin{equation}
i\frac{\partial \Psi }{\partial t}=-\frac{1}{2}\left( \frac{\partial
^{2}\Psi }{\partial r^{2}}+\frac{1}{r}\frac{\partial \Psi }{\partial r}+%
\frac{1}{r^{2}}\frac{\partial ^{2}\Psi }{\partial \phi ^{2}}\right) -\frac{1%
}{2}r^{2\gamma }\Psi +g|\Psi |^{2}\Psi ,  \label{2DNLS}
\end{equation}%
Both the 1D and 2D versions produce nontrivial results in the case of $%
\gamma >1$, i.e., with the expulsive potential which is steeper than the
quadratic (\textit{inverted HOP}) one, that corresponds to $\gamma =1$. The
coefficient in front of the potential term in Eq. (\ref{2DNLS}) is fixed as $%
1/2$ by means of scaling$.$ In terms of BEC, the expulsive potential, in its
1D and 2D forms alike, can be induced by a blue-detuned optical beam (which
exerts repulsion on atoms) with the local intensity decaying from the center
to periphery in the transverse plane \cite{opt-beam2,opt-beam3}.
Alternatively, one can use a red-detuned (attractive) beam, with the local
intensity growing from the center to periphery. The simplest realization of
the latter option may be provided by a red-tuned beam carrying internal
vorticity \cite{vortex}.

In fact, the most essential case is the one with $g=0$, which produces
nontrivial results in the framework of linearized equations (\ref{1DNLS})
and (\ref{2DNLS}): instead of intuitively expected strong delocalization of
the wave function under the action of the steep expulsive potential, the
wave functions demonstrate, quite counter-intuitively, an effect of \emph{%
linear self-trapping}. This means that GS and all excited states produced by
Eqs. (\ref{1DNLS}) and (\ref{2DNLS}) with $\gamma >1$ (in particular, 2D
vortex eigenstates which carry angular momentum) are effectively localized
ones, with a convergent value of the integral norm (\ref{N}).

In this connection, it is relevant to mention that examples of localized
\textit{bound states in continuum} (BIC), which are supported by expulsive
potentials that are expected to maintain solely the continuous spectrum
composed of delocalized modes, are known as exceptional solutions existing
under special conditions. The famous example was reported in 1929 by von
Neumann and Wigner (vNW) \cite{1929}, as a exceptional solution of the 3D
Schr\"{o}dinger equation with the expulsive potential, $%
U(r)=r^{-2}-(9/2)r^{4}$. This exact solution, which corresponds to the
energy eigenvalue $E=0$, is%
\begin{equation}
\varphi _{\mathrm{vNW}}(r)=\frac{1}{r^{2}}\sin \left( r^{3}\right) .
\label{BIC}
\end{equation}%
Due to the similarity between the 2D quantum-mechanical Schr\"{o}dinger
equation and the equation in classical optics which governs the paraxial
propagation of light in bulk waveguides, much interest was attracted to BIC
in various optical setups, which may be employed in various applications
\cite{BIC1,BIC8,BIC10}. In this context, the normalizable solutions of Eqs. (%
\ref{1DNLS}) and (\ref{2DNLS}) with $\gamma >1$ may be considered as
BIC-like states which form the continuous spectrum.

The nonlinear term in Eqs. (\ref{1DNLS}) and (\ref{2DNLS}) produces
nontrivial results in the case of $g<0$, i.e., self-focusing nonlinearity
(opposite to the self-defocusing sign of the cubic nonlinearity in Eq. (\ref%
{GPE})). The result may be spontaneous breaking of symmetry of the
eigenstates produced by the linear versions of these equations \cite%
{Academia}.

The present \textit{perspective} article aims to briefly highlight basic
counter-intuitive findings produced by the work with Eqs. (\ref{GPE}) and (%
\ref{1DNLS}), (\ref{2DNLS}), which include the potentials with the
singularity at $r\rightarrow 0$ and $r\rightarrow \infty $, respectively.
The results for the QC\ suppression by the cubic self-repulsive term in Eq. (%
\ref{GPE}) were reported in some detail in the course of the last 15 years
\cite{HS1,HS2,HS3,GEA,Viskol,HS4}, and they were briefly reviewed in Ref.
\cite{CondMatt}, therefore they are recapitulated in a concise form in
section 2. The most essential part of the article is section 3, which
briefly summarizes very recent findings \cite{Academia} for the model based
on Eqs. (\ref{1DNLS}) and (\ref{2DNLS}). These results surveyed in the
article demonstrated a possibility of producing unexpectedly novel findings
in the framework of the well-established quantum theory, and suggest
possibilities for further developments, especially in nonlinear quantum
models.

\section{Recapitulation of the suppression of the quantum collapse (QC) in
the case of the singular\ inverse-square potential (ISP)}

\subsection{The ground state (GS)}

The GS wave function of Eq. (\ref{GPE}), with chemical potential $\mu <0$,
is looked for as
\begin{equation}
\Psi \left( r,t\right) =\exp \left( -i\mu t\right) \varphi (r),
\label{psiphi}
\end{equation}%
with real function $\varphi (r)$ satisfying the equation%
\begin{equation}
\mu \varphi =-\frac{1}{2}\left( \frac{d^{2}\varphi }{dr^{2}}+\frac{2}{r}%
\frac{d\varphi }{dr}+\frac{U_{0}}{r^{2}}\right) \varphi +\varphi ^{3}.
\label{stationary}
\end{equation}%
In the linear limit (omitting the cubic term), Eq. (\ref{stationary})) gives
rise to two exact solutions with eigenvalue $\mu =0$:
\begin{equation}
\varphi (r)=\varphi _{0}r^{-\sigma _{\pm }},\sigma _{\pm }\equiv \frac{1}{2}%
\pm \sqrt{\frac{1}{4}-U_{0}}~.  \label{exact}
\end{equation}%
The solutions are meaningful if powers $\sigma _{\mp }$ are real in Eq. (\ref%
{exact}), i.e., precisely under condition (\ref{1/4}) ($U_{0}<1/4$). The
non-existence of solution (\ref{exact}) at $U_{0}>1/4$ implies the onset of
QC in the framework of the linear Schr\"{o}dinger equation (\ref{Schr}).
Note that the 2D version of the linear equation also gives rise to two exact
solutions $\varphi (r)=\varphi _{0}r^{-\sigma _{\pm }}~$(cf. Eq. (\ref{exact}%
)), but with $\sigma _{\pm }=\pm \sqrt{-U_{0}}$, hence in the 2D case
arbitrarily small values of the ISP strength, $U_{0}>0$, always lead to QC.

To address the nonlinear stationary equation (\ref{stationary}), it is
convenient to use the substitution
\begin{equation}
\varphi (r)=r^{-1}\chi (r),  \label{phichi}
\end{equation}%
which leads to the equation
\begin{equation}
\mu \chi =-\frac{1}{2}\left( \frac{d^{2}\chi }{dr^{2}}+\frac{U_{0}}{r^{2}}%
\chi \right) +\frac{\chi ^{3}}{r^{2}}.  \label{chi}
\end{equation}%
It is easy to construct asymptotic forms of bound-state solutions to Eq. (%
\ref{chi}) at $r\rightarrow 0$ and $r\rightarrow \infty $. For $r\rightarrow
0$, it is%
\begin{equation}
\chi (r)=\sqrt{U_{0}/2}+\chi _{1}r^{s/2}+...,~s\equiv 1+\sqrt{1+8U_{0}},
\label{r=0}
\end{equation}%
where $\chi _{1}$ is an arbitrary constant \cite{HS1}. At $r\rightarrow
\infty $ and $\mu <0$ (the bound states do not exists with $\mu >0$) the
asymptotic solution is
\begin{equation}
\chi =\chi _{0}\exp \left( -\sqrt{-2\mu }r\right) ,  \label{r=infty}
\end{equation}%
where $\chi _{0}$ is another arbitrary constant. A global analytical
approximation for $\Psi (r)$ (see Eq. (\ref{psiphi})) can be introduced as
an interpolation between the two asymptotic forms:{\
\begin{equation}
\Psi _{\mathrm{interp}}(r,t)=\sqrt{\frac{U_{0}}{2}}e^{-i\mu t}r^{-1}e^{-%
\sqrt{-2\mu }r}.~  \label{inter}
\end{equation}%
The singularity of wave function (\ref{inter}) at }$r\rightarrow 0$, $\Psi
\sim r^{-1}$, is acceptable, as the corresponding norm integral (\ref{N})
converges for all values of the ISP strength $U_{0}$. In fact, it means that
the cubic nonlinear term in Eq. (\ref{GPE}) completely suppresses the QC for
all values of $U_{0}$, thus eliminating the critical value $U_{0}=1/4$, see
Eq. (\ref{1/4}), above which the corresponding linear Schr\"{o}dinger
equation has no GS.

The family of the GS solutions produced by GPE is characterized by
dependence $\mu (N)$. In particular, the interpolated expression (\ref{inter}%
) yields%
\begin{equation}
\mu =-\frac{1}{2}\left( \frac{\pi U_{0}}{N}\right) ^{2}  \label{mu}
\end{equation}%
\cite{HS1}. The scaling relation $\mu \sim N^{-2}$, demonstrated by Eq. (\ref%
{mu}), is a straightforward exact property\emph{\ }of Eq. (\ref{stationary}).

\begin{figure}[t]
\begin{center}
\includegraphics[height=4.cm]{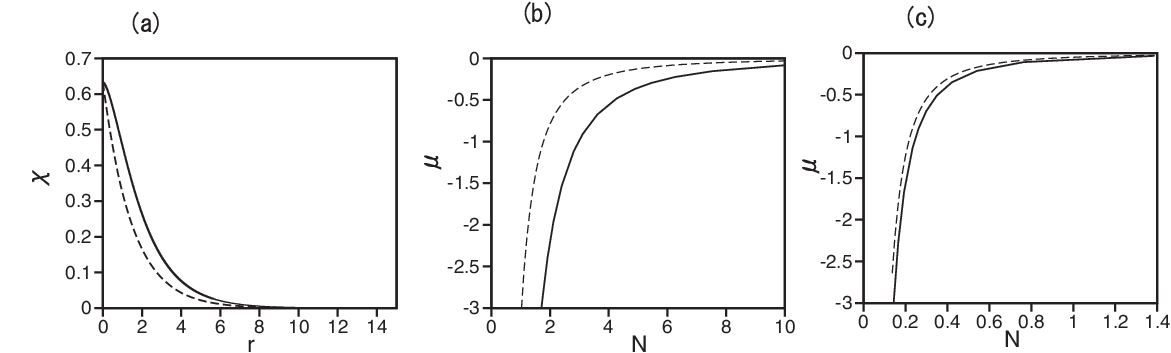}
\end{center}
\caption{ (a) An example of the wave function $\protect\chi (r)$, see Eq. (%
\protect\ref{phichi}), of the three-dimensional GS, produced by the
numerical solution of Eq. (\protect\ref{chi}), for $U_{0}=0.8$ and $\protect%
\mu =-0.225$, as per Ref. \protect\cite{HS1}. Panels (b) and (c) represent
GS\ families by means of dependences $\protect\mu (N)$, for the ISP\
strength $U_{0}=0.8$ and $0.1$, respectively. These svalues are,
respectively, higher and lower than the critical one, $1/4$ (see Eq. (%
\protect\ref{1/4})) for the corresponding linear Schr\"{o}dinger equation (%
\protect\ref{Schr}). Solid and dashed curves represent, respectively, the
numerical solution of Eq. (\protect\ref{chi}), and the respective
approximation based on Eqs. (\protect\ref{inter}) and (\protect\ref{mu}).}
\label{fig1}
\end{figure}

A typical example of the GS, produced by a numerical solution of Eq. (\ref%
{chi}), along with its approximate counterpart (\ref{inter}), is presented
in Fig. \ref{fig1}(a) for $U_{0}=0.8$, which is essentially \emph{larger}
than the critical value $1/4$ (see Eq. \ref{1/4}), above which the linear
Schr\"{o}dinger equation (\ref{Schr}) has no GS. For two values of the
attraction strength, $U_{0}=0.8$ and $0.1$ (larger and smaller than the
critical value $1/4$, respectively), Figs. \ref{fig1}(b) and (c) represent
families of the GS\ states, by means of the corresponding dependences $\mu
(N)$, along with approximation (\ref{mu}). Thus, the self-repulsive cubic
term in Eq. (\ref{GPE}) completely suppresses the QC induced by the singular
ISP in the 3D, space and creates the GS when the QC does not admit its
existence in the framework of the linear Schr\"{o}dinger equation.

Further numerical analysis, based on systematic simulations of the perturbed
evolution of the bound states in the framework of Eq. (\ref{GPE})
corroborate that the entire GS family is stable. \cite{HS1}. This conclusion
agrees with the prediction of the \textit{anti-Vakhitov-Kolokolov} (anti-VK)
criterion, which states that the positive slope of the dependence $\mu (N)$,
i.e., $d\mu /dN>0$, is a necessary stability condition for bound states
supported by self-repulsive nonlinearities \cite{anti}. The VK criterion per
se, $d\mu /dN<0$, is the necessary condition for the stability in
self-attractive systems \cite{VK,Berge'}) .

\subsection{2D and 3D vortex states carrying the angular momentum}

The 2D version of norm (\ref{N}) of the wave function with the asymptotic
form $\varphi (r)\approx \sqrt{U_{0}/2}r^{-1}$ at $r\rightarrow 0$, which is
produced by Eqs. (\ref{phichi}) and (\ref{r=0}), logarithmically diverges at
small $r$, unlike the convergent 3D norm. Therefore, the cubic
self-repulsive term is not strong enough to suppress the 2D collapse. In
this connection, it is relevant to mention that the GPE may include the
quintic self-repulsive term which represents the effect of three-body
collisions \cite{3-body1,3-body2}.

In the scaled form, the 2D equation with the quintic term is written, in the
polar coordinates, $\left( r,\phi \right) $ as
\begin{equation}
i\frac{\partial \Psi _{\mathrm{2D}}}{\partial t}=-\frac{1}{2}\left( \frac{%
\partial ^{2}}{\partial r^{2}}+\frac{1}{r}\frac{\partial }{\partial r}+\frac{%
1}{r^{2}}\frac{\partial ^{2}}{\partial \phi ^{2}}+\frac{U_{0}}{r^{2}}\right)
\Psi _{\mathrm{2D}}+\left\vert \Psi _{\mathrm{2D}}\right\vert ^{4}\Psi _{%
\mathrm{2D}}.  \label{psi2d}
\end{equation}%
Stationary solutions to Eq. (\ref{psi2d}), with chemical potential $\mu $,
are looked for as%
\begin{equation}
\Psi _{\mathrm{2D}}\left( r,\phi ,t\right) =e^{-i\mu t+im\phi }r^{-1/2}\chi
_{\mathrm{2D}}(r),  \label{psichi2D}
\end{equation}%
where integer $m$ is the azimuthal (magnetic) quantum number, alias the
vorticity, and real function $\chi _{\mathrm{2D}}(r)$ satisfies the equation%
\begin{equation}
\mu \chi _{\mathrm{2D}}=-\frac{1}{2}\left[ \frac{d^{2}}{dr^{2}}+\left(
U_{m}^{\mathrm{(2D)}}+\frac{1}{4}\right) r^{-2}\right] \chi _{\mathrm{2D}%
}+r^{-2}\chi _{\mathrm{2D}}^{5},  \label{chi2D}
\end{equation}%
with $U_{m}^{\mathrm{(2D)}}\equiv U_{0}-m^{2}$.

The expansion of the solution to Eq. (\ref{chi2D}) at $r\rightarrow 0$
yields
\begin{equation}
\chi _{\mathrm{2D}}=\left[ \frac{1}{2}\left( U_{m}^{\mathrm{(2D)}}+\frac{1}{4%
}\right) \right] ^{1/4}+\chi _{1}r^{s},  \label{r=0-2D}
\end{equation}%
where $s=\left( 1/2\right) \left( 1+\sqrt{5+16U_{m}^{\mathrm{(2D)}}}\right) $%
, and $\chi _{1}$ is an arbitrary constant, cf. Eq. (\ref{r=0}) \cite{HS1}.
The solution with the convergent norm represents the suppression of the
two-dimensional QC and the creation of the GS ($m=0$) or vortex states (with
$m\geq 1$) by the quintic self-repulsion. Note the drastic difference of
these vortex modes with their \emph{weakly singular} asymptotic form at $%
r\rightarrow 0$, \textit{viz}., $\Psi _{\mathrm{2D}}\left( r,t\right) \sim
e^{-i\mu t+im\phi }r^{-1/2}$, according to Eq. (\ref{psichi2D}), from the
usual vortex wave functions, which have the standard asymptotic form $%
\varphi (r)\sim e^{-i\mu t+im\phi }r^{|m|}$ \cite{2D-review}

\begin{figure}[t]
\begin{center}
\includegraphics[height=4cm]{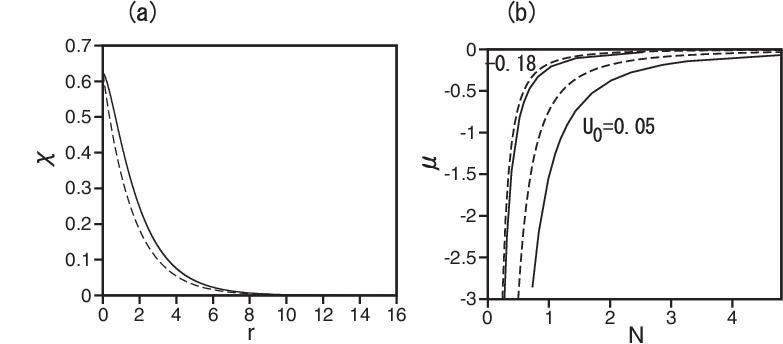}
\end{center}
\caption{(a) The radial profile of the two-dimensional\ GS stabilized by the
quintic self-repulsive term, for $U_{0}=0.05$ and $\protect\mu =-0.1867$.
(b) Curves $\protect\mu (N)$ for the GS with $U_{0}=-0.18$ and $0.05$. The
numerical results and the corresponding approximation (\protect\ref{inter2D}%
) are plotted by the continuous and dashed curves, respectively. The figure
is borrowed from Ref. \protect\cite{HS1}.}
\label{fig2}
\end{figure}

Combining the 2D asymptotic form (\ref{r=0-2D}), valid at $r\rightarrow 0$,
and the obvious approximation valid at $r\rightarrow \infty $, $\chi _{%
\mathrm{2D}}\approx \chi _{0}\exp \left( -\sqrt{-2\mu }r\right) $, one can
derive an interpolating expression, and the dependence $\mu (N)$ following
from it, cf. Eqs. (\ref{inter}) and (\ref{mu}) in the 3D case:
\begin{eqnarray}
\left( \Psi _{\mathrm{2D}}\right) _{\mathrm{interp}}\left( r,t\right)  &=&%
\left[ \frac{1}{2}\left( U_{m}^{\mathrm{(2D)}}+\frac{1}{4}\right) \right]
^{1/4}e^{-i\mu t+im\phi }r^{-1/2}e^{-\sqrt{-2\mu }r},  \notag \\
\mu  &=&-\left( U_{m}^{\mathrm{(2D)}}+\frac{1}{4}\right) \left( \frac{\pi }{%
2N}\right) ^{2},  \label{inter2D}
\end{eqnarray}%
An example of the numerically found 2D\ stable GS, and the respective curves
$\mu (N)$ are plotted, along with approximation (\ref{inter2D}), in Fig.
\ref{fig2} (referring to $m=0$, although $m$ actually makes no difference in the
plots). The $\mu (N)$ curves are shown for both signs of the central
potential, $U_{0}=-0.18$ and $U_{0}=0.05$ (note another \emph{%
counter-intuitive finding}: according to Eqs. (\ref{chi2D}) and (\ref{r=0-2D}%
), the 2D localized GS exists even under the action of the weakly repulsive
ISP, with $-1/4<U_{0}<0$ \cite{HS1} \ -- in particular, at $U_{0}=-0.18$ in
Fig. \ref{fig2}(b)). Simulations of the perturbed evolution in the framework
of Eq. (\ref{psi2d}) confirm stability of the GS families.

The three-dimensional GPE (\ref{GPE}) also produces eigenstates carrying the
angular momentum. To this end, the solution with chemical potential $\mu <0$
and vorticity, which is again represented by the azimuthal (magnetic)
quantum number $m$, are looked for as
\begin{equation}
\Psi \left( r,\theta ,\phi \right) =\exp \left( -i\mu t+im\phi \right)
u^{(m)}\left( r,\theta \right) ,  \label{psi-um}
\end{equation}%
in terms of the spherical coordinates $\left( r,\theta ,\phi \right) $ (cf.
Eq. (\ref{psichi2D})). The substitution of ansatz (\ref{psi-um}) in Eq. (\ref%
{GPE}) produces the stationary equation for real function $u^{(m)}\left(
r,\theta \right) $:%
\begin{gather}
\mu u^{(m)}=-\frac{1}{2}\left[ \frac{\partial ^{2}}{\partial r^{2}}+\frac{2}{%
r}\frac{\partial }{\partial r}\right.  \notag \\
\left. +\frac{1}{r^{2}}\left( \frac{\partial ^{2}}{\partial \theta ^{2}}%
+\cot \theta \cdot \frac{\partial }{\partial \theta }+\left( U_{0}-\frac{%
m^{2}}{\sin ^{2}\theta }\right) \right) \right] u^{(m)}+\left(
u^{(m)}\right) ^{3}.  \label{u2}
\end{gather}%
Following the pattern of Eq. (\ref{phichi}), the substitution%
\begin{equation}
u^{(m)}\left( r,\theta \right) \equiv r^{-1}V^{(m)}\left( r,\theta \right)
\label{uv2}
\end{equation}%
casts Eq. (\ref{u2}) into an equation for the non-singular function $%
V^{(m)}\left( r,\theta \right) $, cf. Eq. (\ref{chi}):
\begin{equation}
2\mu r^{2}V^{(m)}=-r^{2}\frac{\partial ^{2}V^{(m)}}{\partial r^{2}}-\left(
\frac{\partial ^{2}}{\partial \theta ^{2}}+\cot \theta \cdot \frac{\partial
}{\partial \theta }+\left( U_{0}-\frac{m^{2}}{\sin ^{2}\theta }\right)
\right) V^{(m)}+2\left( V^{(m)}\right) ^{3}.  \label{V}
\end{equation}%
Crucially important is the consideration of the limit form of Eq. (\ref{V})
for $r\rightarrow 0$. While in the isotropic case considered above it
amounts to the simple result (\ref{r=0}), in the present case it leads to a
differential equation for function $V^{(m)}\left( r=0,\theta \right) \equiv
v^{(m)}(\theta )$:
\begin{equation}
\frac{d^{2}v^{(m)}}{d\theta ^{2}}+\cot \theta \cdot \frac{dv^{(m)}}{d\theta }%
+\left( U_{0}-\frac{m^{2}}{\sin ^{2}\theta }\right) v^{(m)}=2\left(
v^{(m)}\right) ^{3},  \label{v}
\end{equation}%
which should be solved in the region of $0\leq \theta \leq \pi $. Further
analysis \cite{HS4} demonstrates that the solutions for the vortex states
are categorized by the two usual quantum numbers, \textit{viz}., the orbital
and magnetic ones, $l\geq 1$ and $m\gtrless 0$, subject to the standard
constraint, $|m|\leq l$. The solutions exists if the ISP strength $U_{0}$
exceeds a threshold value,%
\begin{equation}
U_{0}\geq \left( U_{0}^{(l)}\right) _{\mathrm{thr}}=l(l+1),  \label{thr}
\end{equation}%
which does not depend om $m$. Further results may be briefly summarized as
follows \cite{HS4}.

For \emph{sectoral states}, which correspond to $m=l$, the eigenfunction at
the threshold, produced by the linearized version of Eq. (\ref{v}), is

\begin{equation}
v^{(m=l)}(\theta )=v_{0}^{(m=l)}\left( \sin \theta \right) ^{m},  \label{v0}
\end{equation}%
with an infinitesimal amplitude $v_{0}^{(m)}$. Slightly above the threshold,
i.e., at $0<U_{0}-l(l+1)\ll 1$, small finite values of the amplitude are
\begin{equation}
v_{0}^{(m=l=1)}\approx \sqrt{\frac{2}{3}\left( U_{0}-2\right) },  \label{S=1}
\end{equation}%
\begin{equation}
v_{0}^{(m=l=2)}\approx \sqrt{\frac{8}{15}\left( U_{0}-6\right) },
\label{S=2}
\end{equation}%
Examples of the \emph{zonal modes}, which correspond to $m=0$, are%
\begin{equation}
v^{(l=1,m=0)}(\theta )=v_{0}^{(l=1,m=0)}\cos \theta ,  \label{cos}
\end{equation}%
\begin{equation}
v^{(l=2,m=0)}(\theta )=v_{0}^{(l=2,m=0)}\left( 1-3\cos ^{2}\theta \right) ,
\label{cos^2}
\end{equation}%
with the small amplitude $v_{0}^{(l=1,m=0)}$ given by the same expressions (%
\ref{S=1}) as above, while%
\begin{equation}
v_{0}^{(l=2,m=0)}\approx \sqrt{\frac{4}{11}\left( U_{0}-6\right) },
\label{4/11}
\end{equation}%
The set of $l=2$ and $m=1$ corresponds to a \emph{tesseral mode},%
\begin{equation}
v^{(l=2,m=1)}(\theta )=v_{0}^{(l=2,m=1)}\sin \left( 2\theta \right) .
\label{sin2}
\end{equation}%
with the small amplitude
\begin{equation}
v_{0}^{(l=2,m=1)}\approx \sqrt{\frac{2}{3}\left( U_{0}-6\right) }.
\label{quadr}
\end{equation}%
Finally, the entire wave function may be approximated as the interpolation
of the asymptotic forms valid at $r\rightarrow 0$ and $r\rightarrow \infty $%
, cf. Eq. (\ref{inter}):%
\begin{equation*}
\Psi _{\mathrm{interp}}(r,\theta ,\phi ,t)\approx \exp \left( -i\mu t+im\phi
\right) r^{-1}\exp \left( -\sqrt{-2\mu }r\right) v^{(m)}(\theta ).
\end{equation*}

Also considered was the 3D model in which strong external field polarizes
the electric dipole moments of the particles along the $z$ axis. The
respective potential of the interaction with the central electric charge, $U(%
\mathbf{r})=-\left( U_{0}/2\right) r^{-2}\cos \theta $, is cylindrically
symmetric, rather than spherically isotropic. The analysis of GPE (\ref{GPE}%
), with the isotropic attractive potential replaced by this expression, has
demonstrated that such a model also suppresses the QC and maintains stable
bound states. including vorticity-carrying ones \cite{HS2}.

\section{1D and 2D bound states in the strongly expulsive potential}

\subsection{1D bound states}

It is relevant to start the consideration of the stationary states under the
action of the expulsive potential, i.e., one with the singularity at $%
|x|\rightarrow \infty $ (or $r\rightarrow \infty $, in the 2D case), by
looking for eigenstates of Eq. (\ref{1DNLS}), $\Psi (x,t)=\exp \left(
-iEt\right) \varphi (x)$, with eigenvalue (energy/chemical potential) $E$
and real wave function $\varphi (x)$ satisfying the equation
\begin{equation}
E\varphi =-\frac{1}{2}\frac{d^{2}\varphi }{dx^{2}}-\frac{1}{2}r^{2\gamma
}\varphi +g\varphi ^{3}.  \label{1Dpsi}
\end{equation}%
In the framework of an analytical approximation, similar to that which
produces the well-known asymptotic approximation for the Airy function \cite%
{LL}, it is straightforward to construct an asymptotic solution to the
linearized version of Eq. (\ref{1Dpsi}), which is valid for large values of $%
|x|$ and $\gamma \neq 1$ (i.e., for all expulsive potentials, excepts for
the inverted HOP) \cite{Academia}:%
\begin{equation}
\varphi _{\mathrm{asympt}}^{\mathrm{(1D)}}(x;\gamma \neq 1;E)=\varphi
_{0}|x|^{-\gamma /2}\cos \left( \frac{|x|^{\gamma +1}}{\gamma +1}-\chi
_{0}\right) +\frac{E\varphi _{0}}{\gamma -1}|x|^{-3\gamma /2+1}\sin \left(
\frac{|x|^{\gamma +1}}{\gamma +1}-\chi _{0}\right) ,  \label{asympt}
\end{equation}%
In the framework of the asymptotic approximation, amplitude $\varphi _{0}$
and phase shift $\chi _{0}$ are arbitrary constants.

A remarkable property of the wave function (\ref{asympt}) is the convergence
of the respective integral norm, $N_{\mathrm{1D}}=\int_{-\infty }^{+\infty
}\varphi ^{2}(x)dx$, at $|x|\rightarrow \infty $, for $\gamma >1$ (i.e., for
the expulsive potential which is \emph{steeper than inverted HOP}) and all
energies $-\infty <E<+\infty $, which cover the full continuous spectrum.
Thus, a simple but \emph{quite counter-intuitive} conclusion is that any
expulsive potential which is steeper than the quadratic one (inverted HOP),
with $\gamma >1$, produces \emph{normalizable bound states}, which populate
the entire continuous spectrum. This conclusion can be understood by noting
that a classical counterpart of the quantum particle would be rolling down
the steep potential with the rapidly growing acceleration, which gives rise
to rapidly growing oscillations of the phase in the wave function. This
effect induces the effective self-trapping of the eigenstate in the linear
system, manifested by the normalizability of the wave function. The effect
is drastically different from the commonly known mechanism of self-trapping
(creation of solitons) in nonlinear systems \cite{Berge',anti,2D-review}.

For the inverted HOP, with $\gamma =1$, the asymptotic approximation is
irrelevant in the form of Eq. (\ref{asympt}), being replaced by its
modification \cite{Academia}:%
\begin{equation}
\varphi _{\mathrm{asympt}}^{\mathrm{(1D)}}(x;\gamma =1;E)=\varphi
_{0}|x|^{-1/2}\cos \left( \frac{x^{2}}{2}+E\ln \left( \frac{|x|}{l}\right)
\right) ,  \label{gamma=1}
\end{equation}%
with a characteristic scale $l$ of the inner core of the wave function.
Unlike the normalizable bound states with $\gamma >1$, the normalization
integral for wave function (\ref{gamma=1}) slowly diverges at large $|x|$,
\begin{equation}
N_{\mathrm{1D}}\simeq \varphi _{0}^{2}\ln \left( L/l\right) ,  \label{N1D}
\end{equation}%
$2L$ being the size of the integration domain.

\begin{figure}[h]
\begin{center}
\includegraphics[height=4.0cm]{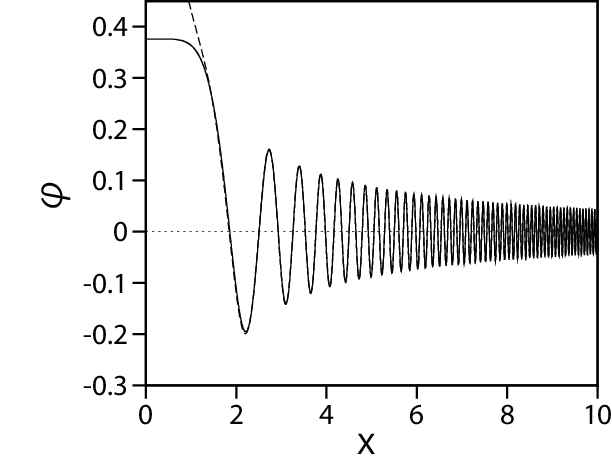}
\end{center}
\caption{The continuous curve represents the numerically found solution of
Eq. (\protect\ref{1Dpsi}) with $\protect\gamma =2$, $g=0$, and $E=0$. The
dashed curve shows the asymptotic approximation (\protect\ref{asympt}) for
the same case, with fitting constants $\protect\chi _{0}=\protect\pi /6$ and
$\protect\varphi _{0}=0.44$. The figure is borrowed from Ref. \protect\cite%
{Academia}. }
\label{fig3}
\end{figure}

The stationary wave function of the 1D eigenstate, produced by a numerical
solution of Eq. (\ref{1Dpsi}) with $\gamma =2$ (the quartic expulsive
potential), $g=0$, and $E=0$, and its analytical counterpart, supplied by
Eq. (\ref{asympt}), are plotted in Fig. \ref{fig3}, which demonstrates high
accuracy of the the analytical approximation.

In addition to the spatially even solution, with $\varphi (-x)=\varphi (x)$,
which represents the GS (alias the fundamental state), the numerical
solution of Eq. (\ref{1Dpsi}) also gives rise to a spatially odd wave
function, with $\varphi (-x)=-\varphi (x)$. It \ represents the first
excited state, and is approximated equally well by Eq. (\ref{asympt}) at
large values of $|x|$ \cite{Academia}. It is feasible that Eq. (\ref{1Dpsi})
gives rise to more sophisticated solutions as well, which represent
higher-order excited states. This issue was not studied, as yet.

\begin{figure}[h]
\begin{center}
\includegraphics[height=4.0cm]{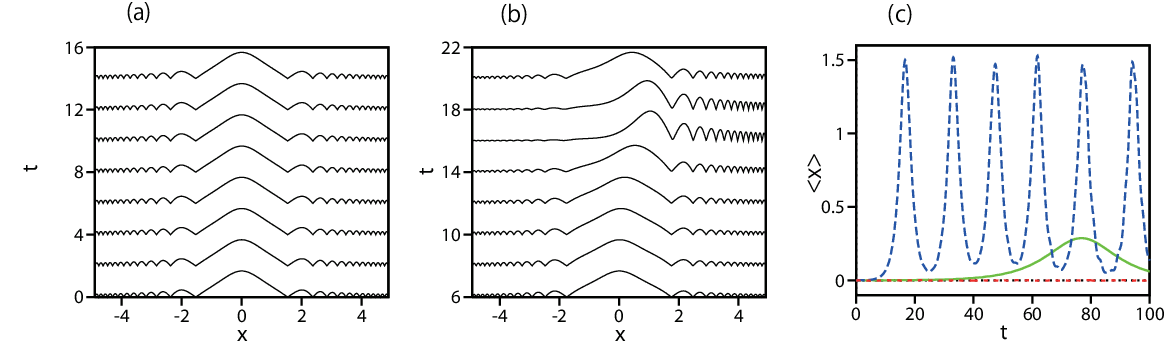}
\end{center}
\caption{(a) The stable evolution of the spatially even GS obtained as the
numerical solution of Eq. (\protect\ref{1DNLS}) with $\protect\gamma =2$ , $%
g=-1$, energy eigenvalue $E=-0.9$, and norm $N_{\mathrm{1D}}=1.73$. Eight
snapshots of $|\protect\psi (x,t)|$ at $t=0,2,\cdots ,14$ are plotted. (b)
Snapshots of $|\protect\psi \left( x,t\right) |$ at $t=6,8,\cdots ,20$
illustrate the case of the spontaneous parity-breaking instability of the
even eigenstate with $E=-1.1$ and $N_{\mathrm{1D}}=3.07$. (c) The
center-of-mass coordinate (\protect\ref{com}) vs. time for the perturbed
evolution of the eigenstates with $E=-0.9$ (the dotted red line), $E=-1,~N_{{%
\mathrm{1}D}}=2.86$ (the solid green line), and $E=-1.1$ (the dashed blue
line).}
\label{fig4}
\end{figure}

If the self-attractive nonlinearity (with $g=-1$) is kept in Eqs. (\ref%
{1DNLS}) and (\ref{1Dpsi}), it affects the stability of the spatially even
GS, as the self-attraction may spontaneously break the symmetry. This
possibility was tested in Ref. \cite{Academia}, running simulations of Eq. (%
\ref{1DNLS}) with input%
\begin{equation}
\Psi (x,t=0)=\varphi _{\mathrm{GS}}(x)+\varepsilon \sin (kx),  \label{t=0}
\end{equation}%
where $\varphi _{\mathrm{GS}}(x)$ is a numerically found spatially even
stationary GS wave function, and $\varepsilon $ is a small amplitude of the
odd perturbation which breaks the GS's spatial parity. Typical results
produced by the simulations for $k=2\pi /L$ and boundary conditions $\Psi
\left( x=\pm L\right) =0$, with $L=4.90$, are displayed in Figs. \ref{fig4}%
(a) and (b) by means of snapshots of instantaneous profiles $|\Psi (x,t)|$,
at times $t=0,2,\cdots ,14$ and $t=6,8,\cdots ,20,$ respectively. In the
former case, the spatially symmetric GS with a smaller norm, $N_{\mathrm{1D}%
}=1.73$ (and $E=-0.9$) remains stable, while larger $N_{\mathrm{1D}}=3.07$
(and $E=-1.1$) leads to the spontaneous breaking of the spatial symmetry in
Fig. \ref{fig4}(b).

The development of the symmetry-breaking instability is illustrated in Fig. %
\ref{fig4}(c) by motion of the solution's center of mass,%
\begin{equation}
\langle x\rangle =\frac{1}{N_{\mathrm{1D}}}\int_{-L}^{+L}x|\Psi (x)|^{2}dx,
\label{com}
\end{equation}%
for $E=-0.9$ (the dotted red line, which corresponds to the stable evolution
in Fig. \ref{fig4}(a)), $E=-1.1$ (the dashed blue line, which corresponds to
the instability in Fig. \ref{fig4}(b)), and also for the intermediate case,
with $E=-1.0$ and $N_{\mathrm{1D}}=2.86$ (the solid green line, which
corresponds to the instability threshold). Detailed studies of the stability
of nonlinear eigenstates remains a subject for subsequent work.

\subsection{2D fundamental and vortex bound states}

2D eigenstates of Eq. (\ref{2DNLS}) with integer vorticity $S$ (the magnetic
quantum number, in terms of atomic physics) and eigenvalue $E$ (alias the
chemical potential, in terms of BEC), are produced by the substitution
\begin{equation}
\Psi =\exp \left( -iEt+iS\theta \right) \varphi (r),  \label{psiphi2D}
\end{equation}%
where real function $\varphi (r)$ obeys the equation%
\begin{equation}
E\varphi =-\frac{1}{2}\left( \frac{d^{2}\varphi }{dr^{2}}+\frac{1}{r}\frac{%
d\varphi }{dr}-\frac{S^{2}}{r^{2}}\varphi \right) -\frac{1}{2}r^{2\gamma
}\varphi +g\varphi ^{3}.  \label{g}
\end{equation}%
The GS (alias the fundamental state) corresponds to $S=0$ in Eqs. (\ref%
{psiphi2D}) and (\ref{g}).

Similar to Eq. (\ref{asympt}) in the 1D case, the asymptotic form of the
solution to Eq. (\ref{g}) at $r\rightarrow \infty $, in the case of $\gamma
\neq 1$, is found as
\begin{equation}
\varphi _{\mathrm{asympt}}^{\mathrm{(2D)}}(r)=\varphi _{0}r^{-(\gamma
+1)/2}\cos \left( \frac{r^{\gamma +1}}{\gamma +1}-\chi _{0}\right) +\frac{%
E\varphi _{0}}{\gamma -1}r^{-\left( 3\gamma -1\right) /2}\sin \left( \frac{%
r^{\gamma +1}}{\gamma +1}-\chi _{0}\right) ,  \label{asympt-2D}
\end{equation}%
with arbitrary constants $\varphi _{0}$ and $\chi _{0}$ \cite{Academia}.

As it follows from Eq. (\ref{asympt-2D}), the 2D norm of the eigenstates, $%
N_{\mathrm{2D}}=2\pi \int_{0}^{\infty }$ $\varphi ^{2}(r)rdr$, converges
under the same condition as in 1D, $\gamma >1$, i.e., if the 2D expulsive
potential is steeper than the inverted HOP. An explanation for this
counter-intuitive conclusion is similar to that proposed above for the 1D
case: a classical particle would be rolling down the steep expulsive
potential (along a spiral trajectory, if the particle carries the angular
momentum, corresponding to $S\geq 1$ in Eq. (\ref{psiphi2D})), with rapid
acceleration. In turn, the acceleration gives rise to rapidly growing
oscillations of the phase of the wave function of the respective quantum
particle. This effect leads to the effective self-trapping of the eigenstate
in the 2D linear system, making its wavefunction normalizable.

The asymptotic solution of the 2D Schr\"{o}dinger equation (\ref{g}) with $%
\gamma =1$ (the inverted-HOP potential) takes a different form,%
\begin{equation}
\varphi _{\mathrm{asympt}}^{\mathrm{(2D)}}(r;\gamma =1)=\varphi
_{0}r^{-1}\cos \left( \frac{r^{2}}{2}+E\ln \left( \frac{r}{l}\right) \right)
,  \label{gamma=1 2D}
\end{equation}%
with\ $l$ being the radius of the inner core, cf. Eq. (\ref{gamma=1}). Also
similar to the 1D counterpart, in the case of $\gamma =1$ the norm of the 2D
eigenstate (\ref{gamma=1 2D}) is weakly divergent, $N_{\mathrm{2D}}\simeq
\pi \varphi _{0}^{2}\ln \left( L/l\right) $ (cf. Eq. (\ref{N1D})), where $L$
is the radius of the solution domain.

Another analytical finding, which is available solely in the 2D case, is the
existence of particular exact solutions to the linear version\ of Eq. (\ref%
{g}), with $g=0$ and $E=0$, if parameter $\gamma $ of the expulsive
potential in Eq. (\ref{2DNLS}) is specifically chosen, for given vorticity $%
S\geq 1$, as%
\begin{equation}
\gamma (S)=2S-1.  \label{gamma(S)}
\end{equation}%
In this case, the exact solution is%
\begin{equation}
\varphi _{\mathrm{exact}}(r;S)=\frac{\varphi _{0}}{r^{S}}\sin \left( \frac{%
r^{2S}}{2S}\right) ,  \label{exact2}
\end{equation}%
with arbitrary amplitude $\varphi _{0}$. The norm of the wave function (\ref%
{exact2}) converges for all $S\geq 2$, and diverges for $S=1$ (i.e., $\gamma
=1$, according to Eq. (\ref{gamma(S)}).

If the cubic term is retained in Eq. (\ref{g}) with $\gamma =1$ (inverted
HOP), it is possible to construct an approximate vortex solution for $E=0$
and $S\neq 1$, neglecting the third harmonic in the relation $\sin ^{3}\phi
=(3/4)\sin \phi -(1/4)\sin \left( 3\phi \right) $. Then, the approximate
solution is
\begin{equation}
\varphi _{\mathrm{approx}}(r)=\sqrt{2\left( 3g\right) ^{-1}\left(
1-S^{2}\right) }r^{-1}\sin \left( \frac{r^{2}}{2}\right) .  \label{squared}
\end{equation}%
In the case of the self-defocusing or focusing nonlinearity, i.e., $g=+1$ or
$g=-1$, this solution exists for $S=0$ or $S\geq 2$, respectively.

\section{Discussion and conclusion}

The aim of this perspective item it to summarize unusual theoretical results
which predict new phenomenology produced by linear and nonlinear quantum
systems which include a potential with the singularity at the center ($%
r\rightarrow 0$) or infinity ($r\rightarrow \infty $). In the former case,
it is the the ISP (inverse-square potential) pulling the quantum particle to
the center, in the 3D or 2D geometry. The main prediction predicted by the
analysis for this case is the stabilization of the QC (quantum collapse) and
creation of the otherwise missing GS (ground state), and higher-order states
carrying angular momentum, by the cubic (in 3D) or quintic (in 2D)
self-repulsive term, which is added to the original linear Schr\"{o}dinger
equation if it is replaced by the GPE (Gross-Pitaevskii equation), in the
framework of the MFA (mean-field approximation), for BEC composed of
particles pulled to the center by the ISP.

This setting calls, first of all, for the experimental realization, as well
as for further theoretical developments. In particular, it may be
interesting to consider the bound states maintained by a set of two mutually
symmetric attractive centers. Another relevant option is to consider the
action of the ISP on 3D and 2D \textit{quantum droplets} (QDs) i.e.,
two-component BEC states intrinsically stabilized \ by the competition of
the weak MFA cubic self-attraction and quartic repulsion induced by the
correction to the MFA induced by quantum fluctuations (the Lee-Huang-Yang
effect \cite{LHY}) \cite{Petrov,AP}. QDs confined by strong or weak HOP
(harmonic-oscillator potential) have been created experimentally in
homonuclear \cite{droplet1,droplet2,droplet3} and heteronuclear \cite%
{droplet4} BEC mixtures. A challenging issue is to analyze the
pulling-to-the-center action of the ISP on a fermionic quantum gas, cf. \cite{Adhikari}.

The very recent analysis of the fundamental and vortex bound states, which
are maintained, in 1D and 2D geometries, by the steep repulsive potential, $%
\sim r^{2\gamma }$, has produced a counter-intuitive conclusion: this
potential, which actually features a singularity at $r\rightarrow \infty $ ,
produces the full spectrum of effectively localized (normalizable) states
for $\gamma >1$, i.e., for the expulsive potentials which are steeper than
the inverted HOP. Such potentials, acting on quantum particles and \ BEC,
can be readily induced by appropriately shaped laser beams. There remain
open questions concerning this setting, especially as concerns effects of
the nonlinearity. In particular, it will be interesting to investigate the
interplay of the critical collapse, induced by the cubic or quintic
self-attraction in the 1D or 2D cases, respectively, with the strong
expulsive potential. Another issue is a possibility of the splitting
instability (alias the azimuthal modulational instability) \cite{2D-review}
of vortex bound states in the 2D system with self-attraction. It is also
relevant to consider the action of the strong expulsive potential on QDs in
the 1D and 2D settings.

\section*{Acknowledgments}

We appreciate valuable collaborations with A. C. Aristotelous, G. E.
Astrakharchik, and E. G. Charalampidis on subjects discussed in this
Perspective article. We also thank D. O'Dell for useful discussions.


\section*{Conflict of interests}

The authors declare no conflict of interest in the context of this paper.


\begin{thebibliography}{99}
\bibitem{LL} Landau, L. D.; Lifshitz, E. M. \textit{Quantum Mechanics:
Nonrelativistic Theory}. Nauka publishers: Moscow, USSR, 1974.

\bibitem{anomaly} Gupta, K. S.; Rajeev, S. G. Renormalization in quantum
mechanics. \emph{Phys. Rev. D} \textbf{1993}, \emph{48}, 5940-5945.

\bibitem{anomaly2} Camblong, H. E.; Epele, L. N.; Fanchiotti, H.; Garc\'{\i}%
a Canal, C. A. Renormalization of the Inverse Square Potential. \emph{Phys.
Rev. Lett}. \textbf{2000}, \emph{85}, 1590.

\bibitem{superselection} \'{A}vila-Aoki, M.;\textit{\ }Cisneros C.; Mart%
\'{\i}nez-y-Romero, R. P.; N\'{u}\~{n}ez-Yepez, H. N.; Salas-Brito, A. L.
Classical and quantum motion in an inverse square potential. \emph{Phys.
Lett. A} \textbf{373}, 418-421 (2009).

\bibitem{Olshanii} Olshanii, M.; Perrin, H.; Lorent, V. Example of a quantum
anomaly in the physics of ultracold gases. \emph{Phys. Rev. Lett.} \textbf{%
2010}, \emph{105}, 095302.

\bibitem{HS1} Sakaguchi, H.; Malomed, B. A. Suppression of the
quantum-mechanical collapse by repulsive interactions in a quantum gas.
\emph{Phys. Rev. A} \textbf{2011}, \emph{83}, 013607.

\bibitem{Trib} Tribelsky M. I. Exact solutions to fall of particle to
singular potential: classical versus quantum cases. \emph{Proc. R. Soc. A}
\textbf{2023}, \emph{479}, 20230366.

\bibitem{ODell} Sundaram S;, Burgess C. P.; O'Dell D. H. J. Duality between
the quantum inverted harmonic oscillator and inverse square potentials,
\emph{New J. Phys}. \textbf{2024}, 26, 053023.

\bibitem{anomaly-new} Zang Y.; Gu Y.; Jiang S. Detecting Quantum Anomalies
in Open Systems. \emph{Phys. Rev. Lett}. \textbf{2024}, \emph{133}, 106503.

\bibitem{LiCs} Deiglmayr, J.; Grochola, A.; Repp, M.; M\"{o}rtlbauer, K.; Gl%
\"{u}ck, C.; Lange, J.; Dulieu, O.; Wester, R.; Weidem\"{u}ller M. Formation
of ultracold polar molecules in the rovibrational ground state. \emph{Phys.
Rev. Lett.} \textbf{2008}, 101, 133004.

\bibitem{KRb} Ospelkaus, S.; Ni, K.-K.; Qu\'{e}m\'{e}ner, G.; Neyenhuis, B.;
Wang, D.; de Miranda, M. H. G.; Bohn, J. L.; Ye, J.; Jin, D. S. Controlling
the hyperfine state of rovibronic ground-state polar molecules. \emph{Phys.
Rev. Lett.} \textbf{2010}, \emph{104}, 030402.

\bibitem{BEC} Pitaevskii, L. and Stringari, S. \textit{Bose-Einstein
Condensation}. Clarendon: Oxford, UK, 2003.

\bibitem{HS2} Sakaguchi, H.; Malomed, B. A. Suppression of the quantum
collapse in an anisotropic gas of dipolar bosons. \emph{Phys. Rev. A}
\textbf{2011}, \emph{84}, 033616.

\bibitem{HS3} Sakaguchi, H.; Malomed, B. A. Suppression of the quantum
collapse in binary bosonic gases. \emph{Phys. Rev. A} \textbf{2013}, \emph{88%
}, 043638).

\bibitem{GEA} Astrakharchik, G. E.; Malomed, B. A. Quantum versus mean-field
collapse in a many-body system, \emph{Phys. Rev. A} \textbf{2015}, \emph{92}%
, 043632.

\bibitem{Viskol} Chen, Z.; Malomed, B. A. Singular and regular vortices on
top of a background pulled to the center, \emph{J. Optics} \textbf{2021}
\emph{23}, 074001.

\bibitem{HS4} Sakaguchi, H.; Malomed, B. A. Angular-momentum modes in a
bosonic condensate trapped in the inverse-square potential. \emph{Symmetry}
\textbf{2023}, \emph{15}, 2060.

\bibitem{Academia} Sakaguchi, H.; Malomed, B. A.; Aristotelous, A. C.;
Charalampidis, G. E. The continuous spectrum of bound states in expulsive
potentials: self-trapping in the linear system, \emph{Academia Quantum}
\textbf{2026}, in press.

\bibitem{opt-beam2} Boyer, V.; Godun, R. M.,; Smirne, G.; Cassettari, D.;
Chandrashekar, C. M.; Deb, A. B.; Laczik, Z. J.; Foot ,C. J., Dynamic
manipulation of Bose-Einstein condensates with a spatial light modulator,
\emph{Phys. Rev. A} \textbf{2006}, \emph{73}, 031402.

\bibitem{opt-beam3} He, C.; Shen, Y.; Forbes, A. Towards higher-dimensional
structured light, \emph{Light: Science \& Applications} \textbf{2022}, \emph{%
11}, 205.

\bibitem{vortex} Lian, Y.; Qi, X.; Wang, Y.; Bai, Z.; Wang, Y.; Lu , Z. OAM
beam generation in space and its applications: A review, Optics and Lasers
in Engineering. \textbf{2022}, \emph{151}, 106923.

\bibitem{2D-review} Malomed, B. A. (INVITED) Vortex solitons: Old results
and new perspectives. \emph{Physica D} \textbf{2019}, \emph{399}, 108-137.

\bibitem{1929} von Neumann, J.; Wigner, E. \"{U}ber merkw\"{u}rdige diskrete
Eigenwerte. \emph{Physikalische Zeitschrift} \textbf{1929}, \emph{30},
465-467.

\bibitem{BIC1} C. W. Hsu, Zhen B., Stone A. D., Joannopoulos J. D., and Solja%
\v{c}i\'{c} M., Bound states in the continuum, Nature Rev. Mat. \textbf{9},
16048 (2016).

\bibitem{BIC8} Kang M., Liu T., Chan C. T., and Xiao, M., Applications of
bound states in the continuum in photonics, Nature Rev. Phys.\textbf{\ 5},
659-678 (2023); https://doi.org/10.1038/s42254-023-00642-8.

\bibitem{BIC10} Huang L, Xu L. Powell D. A., Padilla W. J., and
Miroshnichenko A. E., Resonant leaky modes in all-dielectric metasystems:
Fundamentals and applications, Phys. Rep. 1008, 1-66 (2023),

\bibitem{CondMatt} Malomed, B. A. Suppression of quantum-mechanical collapse
in bosonic gases with intrinsic repulsion: a brief review. \emph{Condensed
Matter} \textbf{2018}, \emph{3}, 15.

\bibitem{anti} Sakaguchi, H.; Malomed, B. A. Solitons in combined linear and
nonlinear lattice potentials. \emph{Phys. Rev. A} \textbf{2010}, \emph{81},
013624.

\bibitem{VK} Vakhitov, M.; Kolokolov, A. Stationary solutions of the wave
equation in a medium with nonlinearity saturation. \emph{Radiophys. Quantum
Electron}. \textbf{1973}, \emph{16}, 783-789.

\bibitem{Berge'} Berg\'{e}, L. Wave collapse in physics: principles and
applications to light and plasma waves. \emph{Phys. Rep}. \textbf{1998},
\emph{303}, 259-370.

\bibitem{3-body1} Abdullaev, F. Kh.; Gammal, A., Tomio, L.; Frederico, T.
Stability of trapped Bose-Einstein condensates. \emph{Phys. Rev. A} \textbf{%
200}1, \emph{63}, 043604.

\bibitem{3-body2} Abdullaev, F. Kh.; Salerno, M. Gap-Townes solitons and
localized excitations in low-dimensional Bose-Einstein condensates in
optical lattices. \emph{Phys. Rev. A} \textbf{2005}, \emph{72}, 033617.

\bibitem{LHY} Lee, T. D.; Huang, K.; Yang, C. N. Eigenvalues and
eigenfunctions of a Bose system of hard spheres and its low-temperature
properties. \emph{Phys. Rev}. \textbf{1957}, \emph{106}, 1135-1145.

\bibitem{Petrov} Petrov, D. S. Quantum mechanical stabilization of a
collapsing Bose-Bose mixture.\emph{\ Phys. Rev. Lett}. \textbf{2015}, \emph{%
115}, 155302.

\bibitem{AP} Petrov, D. S.; Astrakharchik, G. E. Ultradilute low-dimensional
liquids, \emph{Phys. Rev. Lett.} \textbf{2016}, \emph{117}, 100401.

\bibitem{droplet1} Cabrera, C. R.; Tanzi, L.; Sanz, J.; Naylor, B.; Thomas,
P.; Cheiney, P.; Tarruell, L. Quantum liquid droplets in a mixture of
Bose-Einstein condensates. \emph{Science} \textbf{2018}, \emph{359}, 301-304.

\bibitem{droplet2} Cheiney, P.; Cabrera, C. R.; Sanz, J.; Naylor, B.; Tanzi,
L.; Tarruell, L. Bright soliton to quantum droplet transition in a mixture
of Bose-Einstein condensates. \emph{Phys. Rev. Lett}. \textbf{2018}, \emph{%
120}, 135301.

\bibitem{droplet3} Semeghini, G.; Ferioli, G.; Masi, L.; Mazzinghi, C.;
Wolswijk, L.; Minardi, F.; Modugno, M.; Modugno, G.; Inguscio, M.; Fattori,
M. Self-bound quantum droplets in atomic mixtures. \emph{Phys. Rev. Lett}.
\textbf{2018}, \emph{120}, 235301.

\bibitem{droplet4} D'Errico, C.; Burchianti, A.; Prevedelli, M.; Salasnich,
L.; Ancilotto, F.; Modugno, M.; Minardi, F.; Fort. C. Observation of quantum
droplets in a heteronuclear bosonic mixture. Phys. Rev. Research \textbf{2019%
}, \emph{1}, 033155.

\bibitem{Adhikari}
Adhikari, S.; Malomed, B. A. Tightly bound gap solitons in a Fermi gas. \emph{Europhys. Lett.}
\textbf{79}, \emph{50003} (2007).
\end{thebibliography}
\end{document}